\documentclass[a4paper,11pt]{article}
\pdfoutput=1

\usepackage{jheppub}

\usepackage{xspace}
\usepackage[countmax]{subfloat}
\usepackage{slashed}
\usepackage{longtable}

\usepackage{booktabs}
\usepackage{comment}

\usepackage{bbm}

\usepackage{graphicx}
\usepackage{dcolumn}
\usepackage{bm}

\usepackage{xspace}
\usepackage[normalem]{ulem}
\usepackage{textcomp}
\usepackage{placeins}
\usepackage{float}

\usepackage{calrsfs}
\DeclareMathAlphabet{\pazocal}{OMS}{zplm}{m}{n}
\newcommand{\disc}{\pazocal{D}}
\newcommand{\gen}{\pazocal{G}}

\newcommand\pT{\ensuremath{p_\mathrm{T}}\xspace}
\usepackage{subfigure}
\usepackage{color}
\definecolor{darkblue}{rgb}{0,0,0.5}
\definecolor{darkred}{rgb}{0.5,0,0}
\definecolor{darkgreen}{rgb}{0,0.5,0}

\DeclareRobustCommand{\MC}{^\mathrm{\,MC}}
\DeclareRobustCommand{\data}{^\mathrm{\,data}}
\DeclareRobustCommand{\ProbabilityOf}[2]{ P\big( #1 \vphantom{#1 #2} \big) }
\DeclareRobustCommand{\GivenProbOf}[2]{ \ProbabilityOf{ #1 | #2 }{#2} }
\DeclareRobustCommand{\NumberOf}[2]{n\big( #1 \vphantom{#1 #2} \big) }

\newcommand*{\pythiaeight}{\textsc{Pythia}~v8.23}
\newcommand*{\kt}{\ensuremath{k_\mathrm{t}}\xspace}
\newcommand*{\akt}{anti-\kt\xspace}

\def\log{\text{log}}

\def\be{\begin{equation}}
\def\ee{\end{equation}}

	\title{Unfolding with Generative Adversarial Networks}

	\author[1]{Kaustuv Datta,}
	\emailAdd{kdatta@ethz.ch}
	\affiliation[1]{%
		ETH Z\"{u}rich, 8093 Z\"{u}rich, Switzerland
	}%
	\author[2]{Deepak Kar}
	\emailAdd{deepak.kar@cern.ch}
	\affiliation[2]{%
		School of Physics, University of Witwatersrand, Johannesburg, South Africa
	}%
	\author[2]{and Debarati Roy}
	\emailAdd{debarati.roy@cern.ch}

	\date{\today}
	
	\abstract{		
		Correcting measured detector-level distributions to particle-level is essential to make data usable outside the experimental collaborations. The term unfolding is used to describe this procedure. A new method of unfolding data using a modified Generative Adversarial Network (MSGAN) is presented here.
		Applied to various distributions with widely different shapes, it performs roughly at par with currently used methods. This is a proof-of-principle demonstration of 
		a state-of-the-art machine learning method that can be used to model detector effects well. 		
		}
	
\begin{document}
\maketitle

\section{\label{sec:intro} Introduction}

Measurements are an important part of any collider physics programme, even though the main focus
is on discovering new physics. Measurements not only help to validate Standard Model (SM) predictions
at a new energy regime, but also help to improve the Monte Carlo (MC) event generators to model
SM processes acting as backgrounds to searches.

Measurements are usually performed in a well-defined fiducial phase space, and corrected for detector effects,
in order to make the results independent of the details of the particular experiment. 
Unfolding is the generic term used to describe this \textit{detector-level}
to \textit{particle-level} correction for distributions. In general, this can account for limited acceptance, 
finite resolution, and inefficiencies of the detector, 
as well as bin migrations in the distribution between measured and corrected. Mathematically 
this is an ill-posed problem~\cite{Blobel:2002pu}, as an unique answer can not be assumed, and small changes in
the measured distribution can often cause large changes in the corrected distribution.
Simulated samples from MC event generators are used to perform the unfolding. 
Distributions obtained from the generated events correspond to particle-level (will be referred to as \textit{gen}).
Then the events are passed through a detector simulation programme, mimicking the behaviour
of the intended detector as closely as possible, and the same distributions
obtained using these events correspond to detector-level (will be referred to as \textit{ reco}). 

Different methods have been used for unfolding.
The simplest method is bin-by-bin unfolding, where the ratio of unfolding factor $=$ (value-particle-level/value-detector-level)
is obtained for each bin in the distribution from MC, and then the value-data in that particular bin
is multiplied by the unfolding factor for that bin to obtain the corrected data value in that bin.
This simple method does not account for bin migrations, and for a steeply falling distribution, that is
a significant shortcoming. 
In order to address this, two other methods, Bayesian Iterative Unfolding~\cite{D'Agostini:1994zf} and Singular
Value Decomposition~\cite{Hocker:1995kb} are often used. 

\begin{itemize}
	
	\item Bayesian iteratve unfolding: this method produces a matrix relating the number of events in each bin of the measured data 
	distribution $\NumberOf{R_j\data}{C_i\data}$ with the number in each bin of the corrected distribution 
	$\NumberOf{C_i\data}{R_j\data}$, 
	\begin{equation}
	\NumberOf{C_i\data}{R_j\data} = \sum\limits_j \GivenProbOf{T_i\MC}{R_j\MC}\,\NumberOf{R_j\data}{C_i\data},
	\label{eqn:MakeCorrected}
	\end{equation}
	where $\GivenProbOf{T_i\MC}{R_j\MC}$ is the unfolding matrix, calculated using Bayes' theorem as shown in Eq.~\eqref{eqn:UnfoldingMatrix},
	\begin{equation}
	\GivenProbOf{T_i\MC}{\,R_j\MC} = \frac{ \GivenProbOf{R_j\MC}{\,T_i\MC} \ProbabilityOf{T_i\MC}{R_j\MC} }{ \ProbabilityOf{R_j\MC}{T_i\MC} }.
	\label{eqn:UnfoldingMatrix}
	\end{equation}
	The matrix $\GivenProbOf{R_j\MC}{\,T_i\MC}$ is derived from MC, indicating the probability 
	of an event from bin $i$ of the MC truth distribution being found in bin $j$ of the 
	MC reconstructed distribution.
	These are called smearing matrices (alternatively response or migration matrices).
	The MC generator level distribution is used to give the initial Bayesian prior 
	probability $\ProbabilityOf{T_i\MC}{R_j\MC}$.
	
	To avoid dependence of the correction on the choice of prior the process is iterated,
	with the corrected data distribution produced in each iteration used as the prior for the next.
	Three/four iterations are performed as this gives smallest uncertainties: statistical uncertainties accumulate with more iterations, 
	but using only one iteration would not account for prior-dependence.	
	
	\item Singular value decomposition: this method is based on matrix inversion, but uses a regularisation technique based on
	Singular Value Decomposition (SVD) technique in order to suppress statistical fluctuations.
	
\end{itemize}

In this paper, we introduce a semi-supervised, 
adversarial unfolding procedure based on Generative Adversarial Networks (GANs)~\cite{2014arXiv1406.2661G}.
While machine learning methods have been used for unfolding~\cite{Gagunashvili:2010zw, Glazov:2017vni}, 
here for the first time we demonstrate the applicability of using generative adversarial techniques to unfold distributions 
for realistic experimental observables. To the best of our knowledge, this is also the first use of non-image based GANs for high energy physics applications.

In section \ref{sec:setup}, we review the standard statistical tools for unfolding, and our datasets. 
In section \ref{sec:GAN}, we briefly review the standard GAN framework, and then describe our proposed 
MSGAN (Mean Squared Error GAN) framework for unfolding. We present our results for unfolding with 
the MSGAN in section \ref{sec:res}, and compare its performance to standard methods. 
We conclude in section \ref{sec:out} and discuss other potential applications for the MSGAN framework.

\section{\label{sec:setup} Method}
	
In order to demonstrate the success of the method, five distributions with distinct shapes were chosen, 
and they were constructed such that gen and reco distributions have significant differences. 
For the first four jet substructure observables,
boosted $t\overline{t}$ events decaying hadronically were generated 
using \pythiaeight~\cite{Sjostrand:2007gs, Sjostrand:2014zea}.          
Monash~\cite{Skands:2014pea} tune was used to generate training samples, and a new tune based on
thermodynamical string fragmentation~\cite{Fischer:2016zzs} was used to generate the testing sample. 
The details of individual tunes are irrelevant, the aim was to get two sufficiently different distributions. 
To get a significant shift between the gen and reco distributions, the Rivet v2.5.4~\cite{Buckley:2010ar} smearing machinery was employed. 
The same smearing was used for both training and testing samples.
The observables were constructed using the (constituents of) leading \akt~\cite{Catani:1993hr} jet with a 
radius parameter of $R=1.0$, with a  $\pT > 200$~GeV and $|\eta| < 2.0$, mimicking standard experimental event selection.
The observables chosen were mass divided by transverse momentum (i.e., normalised mass), 
one of the energy correlation functions~\cite{Larkoski:2013eya} (ECF2), the Les Houches angularity~\cite{Gras:2017jty} (LHA), 
and the momentum sharing fraction~\cite{Larkoski:2014wba} ($z_g$).
The $z_g$ has a steeply falling spectrum, while ECF2 has a more gradual falling spectrum, along with a spread out Gaussian peak. 
Normalised mass has a steep Guassian peak, while the LHA is a more spread out Gaussian. 
Finally, missing energy (MET) distribution constructed using the MadGraph v2.60~\cite{Alwall:2014hca} with \pythiaeight, and only \pythiaeight~
both smeared with 
Delphes v3.4.1~\cite{deFavereau:2013fsa} 
(ATLAS parametrisation, with average pileup of 50)
from $Z \rightarrow \nu \nu$ process is also looked at, as MET typically has the most dramatic difference between 
detector and particle level. 
However, the point of this exercise is not to produce the exact experimental distribution, 
rather to show the method works for different classes of distributions encountered in general.

\section{\label{sec:GAN}GANs and Adversarial Unfolding}

GANs have recently been explored in the context of high energy physics with the motivation to complement 
current methods, specifically in the area of fast detector simulations ~\cite{deOliveira:2017pjk,deOliveira:2017rwa,Paganini:2017dwg,Paganini:2017hrr, Musella:2018rdi}. 
In this section we will briefly explore the standard GAN formulation and training procedure, 
and then compare that to the MSGAN proposed here for the purpose of unfolding.

A standard GAN framework poses the problem of learning a data distribution 
as a two-player minimax game between a discriminator $\disc(x;\theta_D)$ 
and a generator $\gen(z;\theta_G)$ that seek to minimize their individual cost. 
Here, $\theta_G$ and $\theta_D$ are the multilayer perceptron (MLP) parameters. 
The goal of this approach is to make the GAN tend towards the Nash equilibrium for the game 
by training the generator to produce samples similar to the data distribution, 
such that the discriminator is eventually maximally confused between true and generated samples. 

The generator takes an input of an $m$-dimensional vector of random values $z$, 
drawn from a latent distribution $p_z(z)\in\mathbb{R}^m$. 
Typically, $z$ is drawn from a random normal or uniform distribution. 
If $n$ is the dimensionality of the sample space, then the objective of the generator is to learn a mapping  
$\gen:\mathbb{R}^m\rightarrow\mathbb{R}^n$ from the latent to sample space. The discriminator 
takes inputs from either the sample space, $x\sim p_{data}$, or the space of synthetic 
samples of the generator, $x\sim p_{fake}$. Thus, the discriminator in this setup learns the 
mapping $\disc:\mathbb{R}^n\rightarrow(0,1)$, whereby it predicts whether the input it receives 
was synthesised by the generator $(0)$ or if it was from the true sample space $(1)$. 
The system attains convergence by updating parameters for the standard GAN value function, 
\begin{align*} \min\limits_{\gen}& \max\limits_{\disc}~V(\disc,~\gen) = \mathbb{E}_{x\sim p_{data}(x)}[\log\disc(x)]+\mathbb{E}_{z\sim p_{z}(z)}[\log(1-\disc(\gen(z)))] \tag{2}, \end{align*}
whereby, ideally, the generator probability distribution will eventually match that of the data.	

The discriminator parameters $\theta_D$ are updated by ascending its stochastic gradient on 
mini-batches of data and generated samples. 
This is essentially a binary classification problem, 
where the discriminator is made to update its weights in each training iteration 
to tell apart true and fake (generated) samples via a binary cross entropy (BCE) objective. 
This is carried out through a weight update on fake samples from the generator $x\sim \gen(z;\theta_G)$,

\begin{equation}
\mathcal{L}_{fake}(x)= \mathcal{L}_{BCE}(\disc(x;\theta_D),0) = - \log (1-\disc(x;\theta_D)),\newline
\end{equation}
\newline where $\mathcal{L}_{BCE}(\hat{x},x)=-x \log(\hat{x})-(1-x)\log(1-\hat{x})$, followed by a second update 
on true samples from $x\sim p_{data}$,

\begin{equation}
\mathcal{L}_{true}(x)= \mathcal{L}_{BCE}(\disc(x;\theta_D),1) = -\log (\disc(x;\theta_D)),
\end{equation}

Then, the generator parameters $\theta_G$ are updated by descending its stochastic gradient 
on mini-batches of noise samples from the prior $p_z(z)$,

\begin{align}
\mathcal{L}_{gen}(z)= & \mathcal{L}_{BCE}(\disc(\gen(z;\theta_G);\theta_D),1)\\ = &- \log (\disc(\gen(z;\theta_G);\theta_D)).
\end{align}
This is a forward pass through the whole GAN framework, where the generator output on latent samples $z$, 
with the BCE target set to 1, is passed through the discriminator. The discriminator weights are usually frozen at this stage, 
and the gradient of the BCE loss is then back-propagated through the generator network to train it to produce 
fake samples that better approximate the true samples. 

There are several difficulties associated with the implementation of this formulation, 
and we refer the reader to \cite{NIPS2016_6125} for a 
detailed review of both the difficulties with training GANs and possible methods to mitigate them. 
With a view to bypass some of the issues related with attaining the Nash Equilibrium for the problem 
of unfolding variables with different shapes of distribution, we propose the MSGAN framework.

\subsection{MSGAN}


The first major difference between the standard GAN and MSGAN is the introduction of 
some supervision into the training of the generator. Between the BCE updates to the discriminator and the generator, 
we introduce a supervised regression step that specifically trains the generator to 
minimize the mean squared error (MSE) loss of the generated samples to data, \footnote{One can similarly use mean absolute error (MAE) loss in this step, however the choice of loss function for the regression will depend on what is most appropriate for a given use-case.}
\begin{align}
\mathcal{L}_{gen}(z)= & \mathcal{L}_{MSE}(\gen(z;\theta_G),x).
\end{align}
This explicit supervised step promotes the generator's ability and complements the subsequent 
unsupervised update to the generator with BCE loss.  Once the generator is trained it learns a 
transformation that encodes the detector effects that shift the reco and gen data. 
This approach with an MSE step is similar to an application to the problem of 
image deconvolution presented in Ref.\cite{8190958}, and in Ref.\cite{Paganini:2017dwg} 
where the minimization of mean absolute error (MAE) of requested and reconstructed 
energies is combined with the adversarial loss of the generator. 

The second difference lies in the utilization of a novel latent space. 
Usually, the latent space $z\sim p_z(z)$ is a random noise vector. 
Here, instead, the latent features are distributed according to the reco values, 
$z\sim p_{reco}$ for the chosen observables to be unfolded. 
The generator $\gen$ then learns the mapping from the reco 
latent space to gen samples, $x\sim p_{gen}$, through this semi-supervised, adversarial setup. 

For unfolding, we note that this approach helps to steadily guide the model towards a 
convergence on the data distribution. Attempts with just the standard BCE update 
on the generator, while utilizing the latent vector $z\sim p_{reco}$, were promising but 
much more prone to fluctuations in the loss that led to divergences in the training.

\subsection{Architecture and training details}
We conduct our tests, the results for which are presented in section \ref{sec:res}, 
with the same fully connected network architecture. We note that better performance can be obtained by individualizing the choice of network architecture for each variable, however that is not the main objective of this study. Instead we seek to show that one can conceivably use a simple, fully-connected network to unfold several different shapes of distributions adequately.

Both the generator and 
discriminator networks receive a single input column. The generator is a relatively simple network with only one hidden layer. It consisted 
of fully-connected layers with $1000\rightarrow250\rightarrow1$ nodes, with the input and hidden layers using the leaky rectified linear unit (LeakyReLU) activation~\cite{Maas13rectifiernonlinearities}, while a linear activation is used for the final layer. The generator is compiled with the MSE loss function, 
using Adam~\cite{DBLP:journals/corr/KingmaB14} optimization with a learning rate of $10^{-4}$.

The discriminator network starts with three fully-connected layers with LeakyReLU 
activations, $1000\rightarrow500\rightarrow250$ nodes, followed by a 
mini-batch discrimination (MBD)~\cite{2016arXiv160603498S} 
layer with 5 one-dimensional kernels, using the recipe of \cite{Paganini:2017dwg}. 
The batch features are appended to the last fully connected layer's output, and passed through a 
$\tanh$ activation to the output node which then uses a sigmoid activation to map samples to the 
space $[0,1]$. The discriminator is compiled with BCE loss, 
using Adam optimization with a higher learning rate ($10^{-3}$) than the generator. 
These two networks are then stacked to build the MSGAN framework. 

Networks were trained to unfold individual variables. 
The training procedure we utilize includes one-sided label smoothing~\cite{2016arXiv160603498S} 
and separate updates for the discriminator on true and fake samples, followed by the supervised 
regression step and the adversarial update to the generator. We use a batch size of 5000 and 
the MSGAN requires between $\sim1000-5000$ epochs to produce acceptable results for closure on 
a validation sub-sample (5\%) of the training data, for each chosen variable. A (binning dependent) stopping 
criterion was used wherein acceptable lower and upper bounds were set on the ratio of fake and true 
gen samples, for a given set of reco samples $z$, in bins around the peak of the data distributions. 
The network's performance was further assessed on the separate test dataset generated with a different tuning. 

All deep learning implementations used Keras 2.1.1 libraries \cite{chollet2015keras}. 
Training and inference for the networks were carried out on an Nvidia GeForce GTX 1060 (Laptop) graphics processing unit. 
The Bayesian and SVD unfolding is performed using the RooUnfold package~\cite{adye_2011}.

\section{\label{sec:res} Results}

\begin{figure*}[!htb]
	\centering
	\subfigure[]{\includegraphics[width=0.48\textwidth]{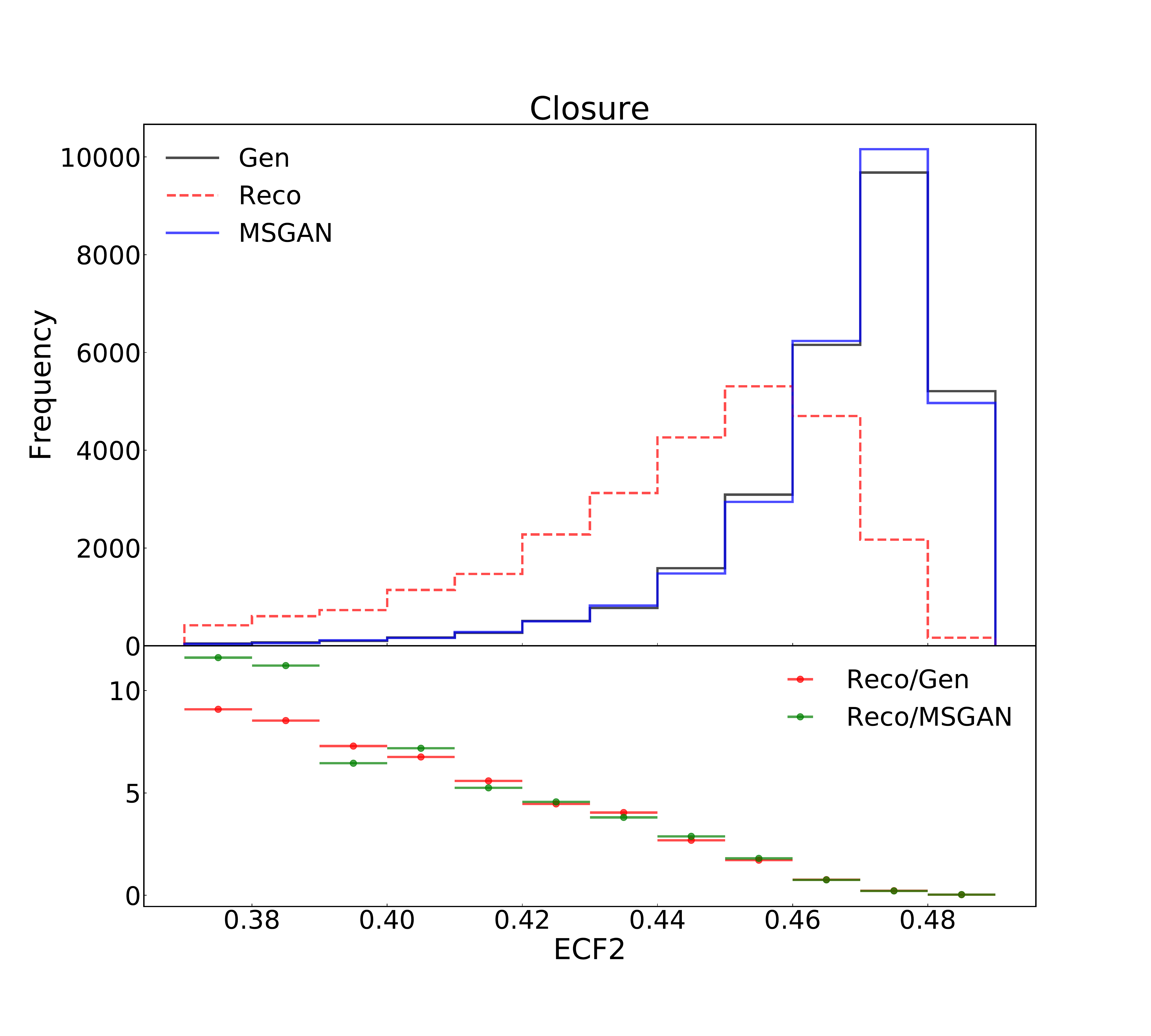}}\quad
	\subfigure[]{\includegraphics[width=0.48\textwidth]{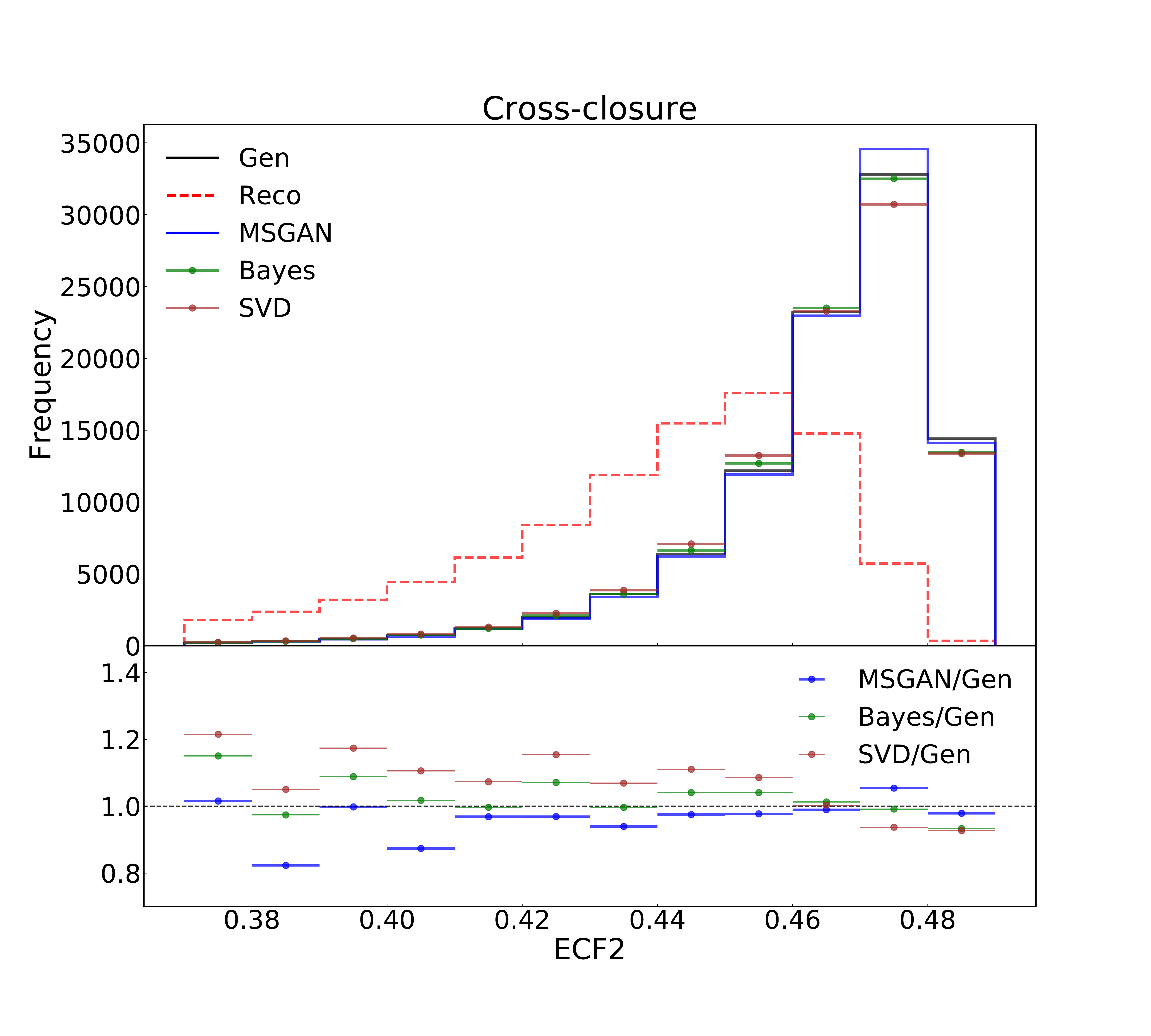}}
	\caption{Energy correlation function unfolding}
	\label{fig:ecf}
\end{figure*}

\begin{figure*}[!htb]
	\centering
	\subfigure[]{\includegraphics[width=0.48\textwidth]{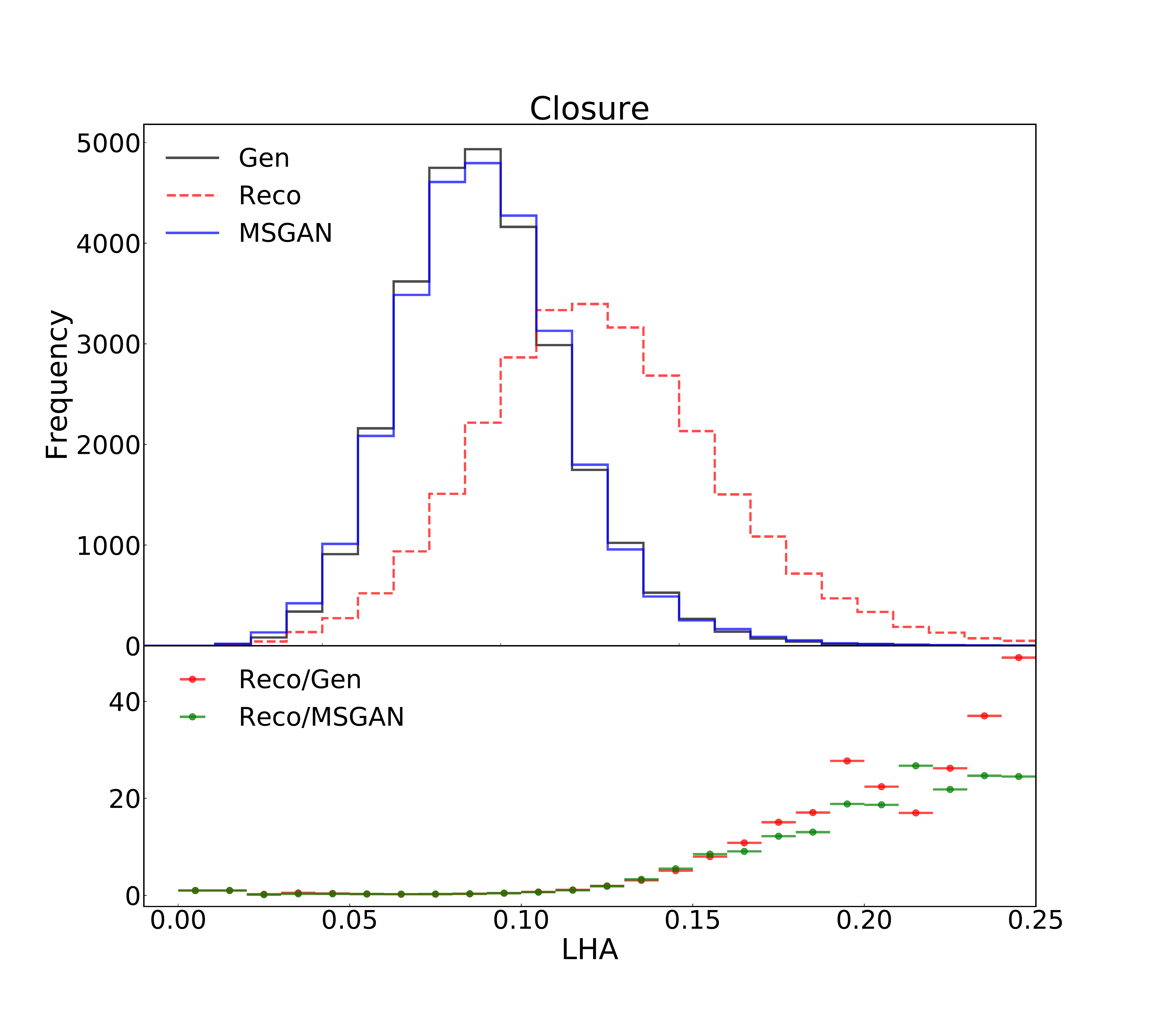}}\quad
	\subfigure[]{\includegraphics[width=0.48\textwidth]{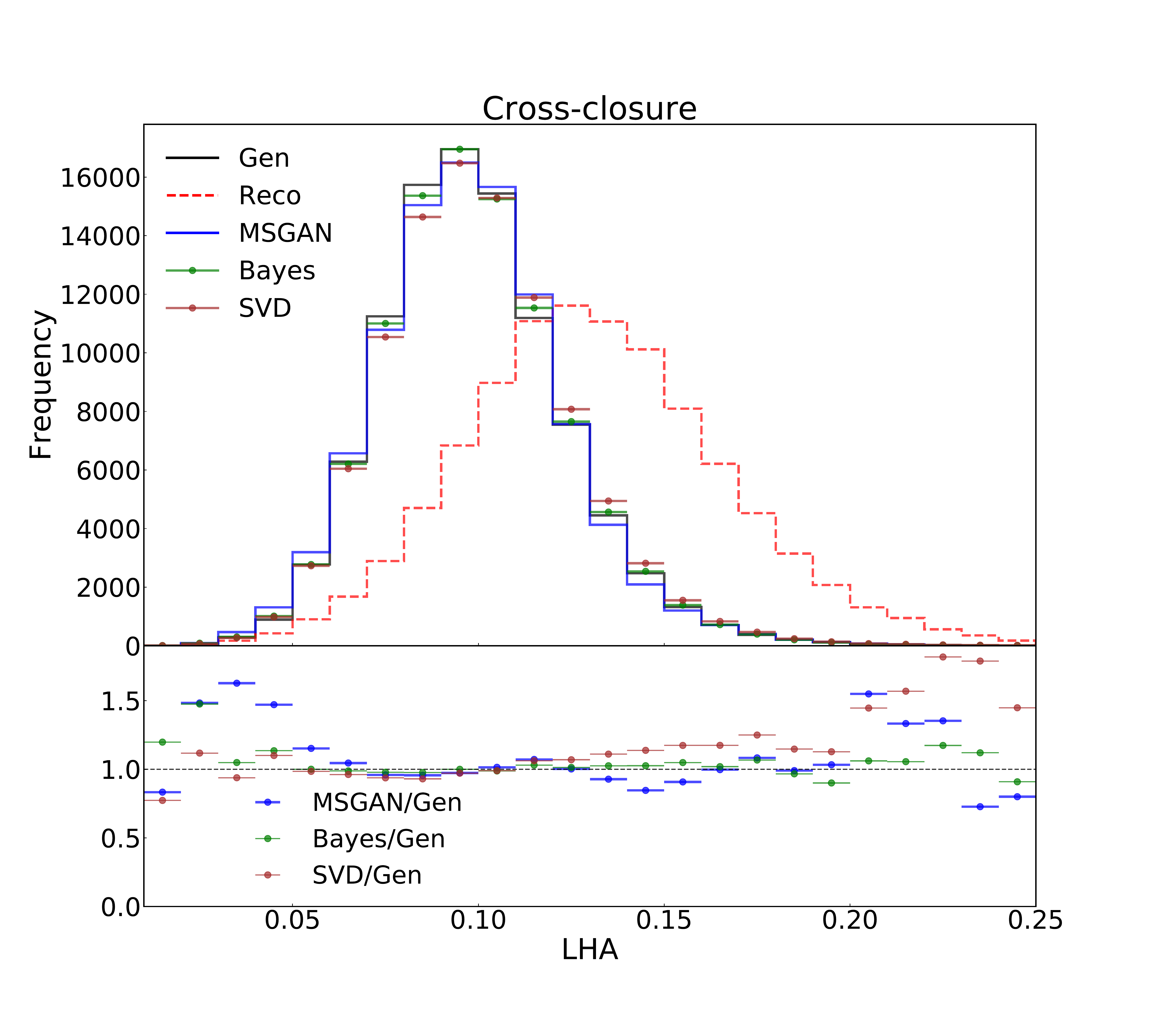}}
	\caption{LHA unfolding}
	\label{fig:lha}
\end{figure*}

\begin{figure*}[!htb]
	\centering
	\subfigure[]{\includegraphics[width=0.48\textwidth]{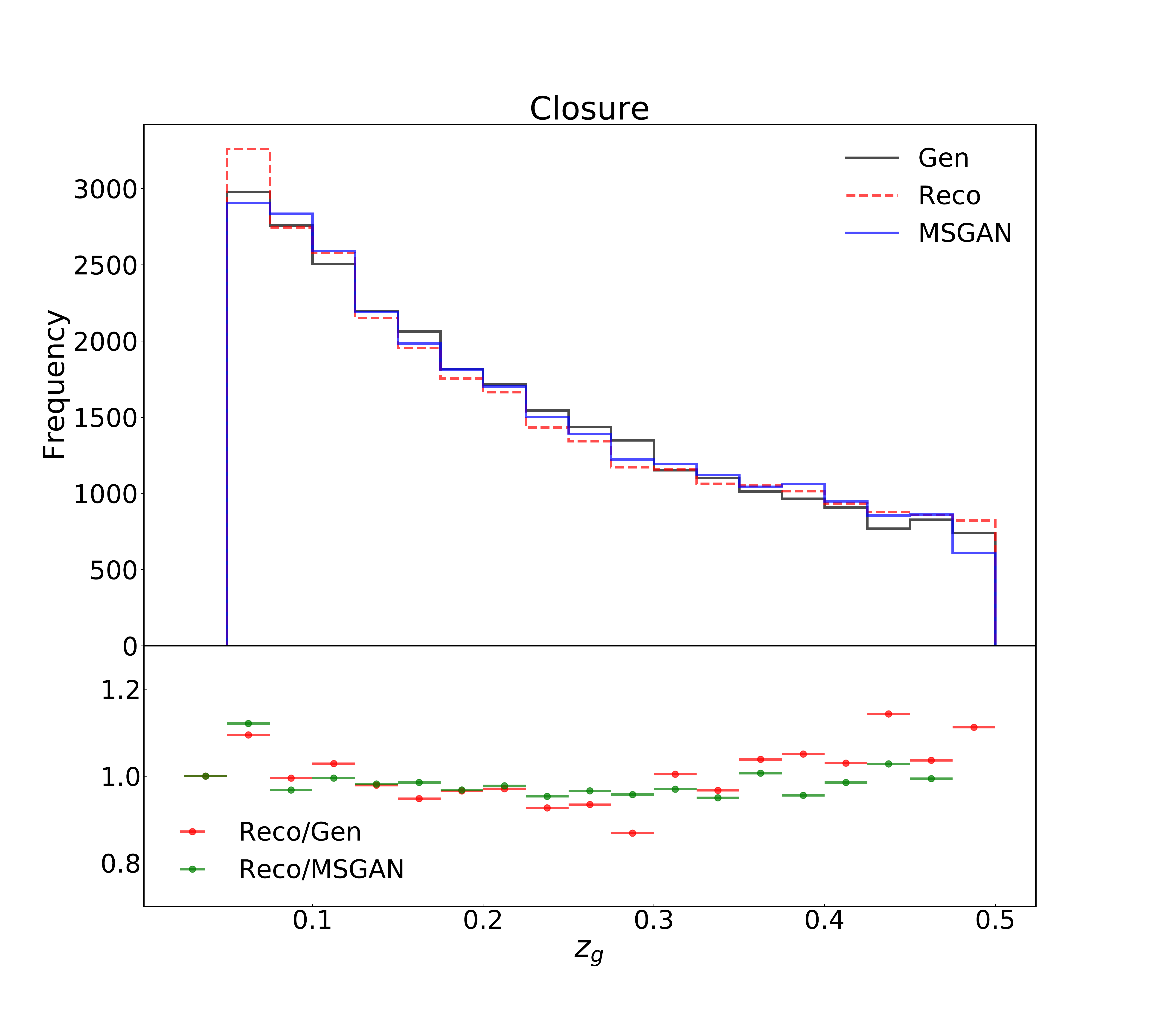}}\quad
	\subfigure[]{\includegraphics[width=0.48\textwidth]{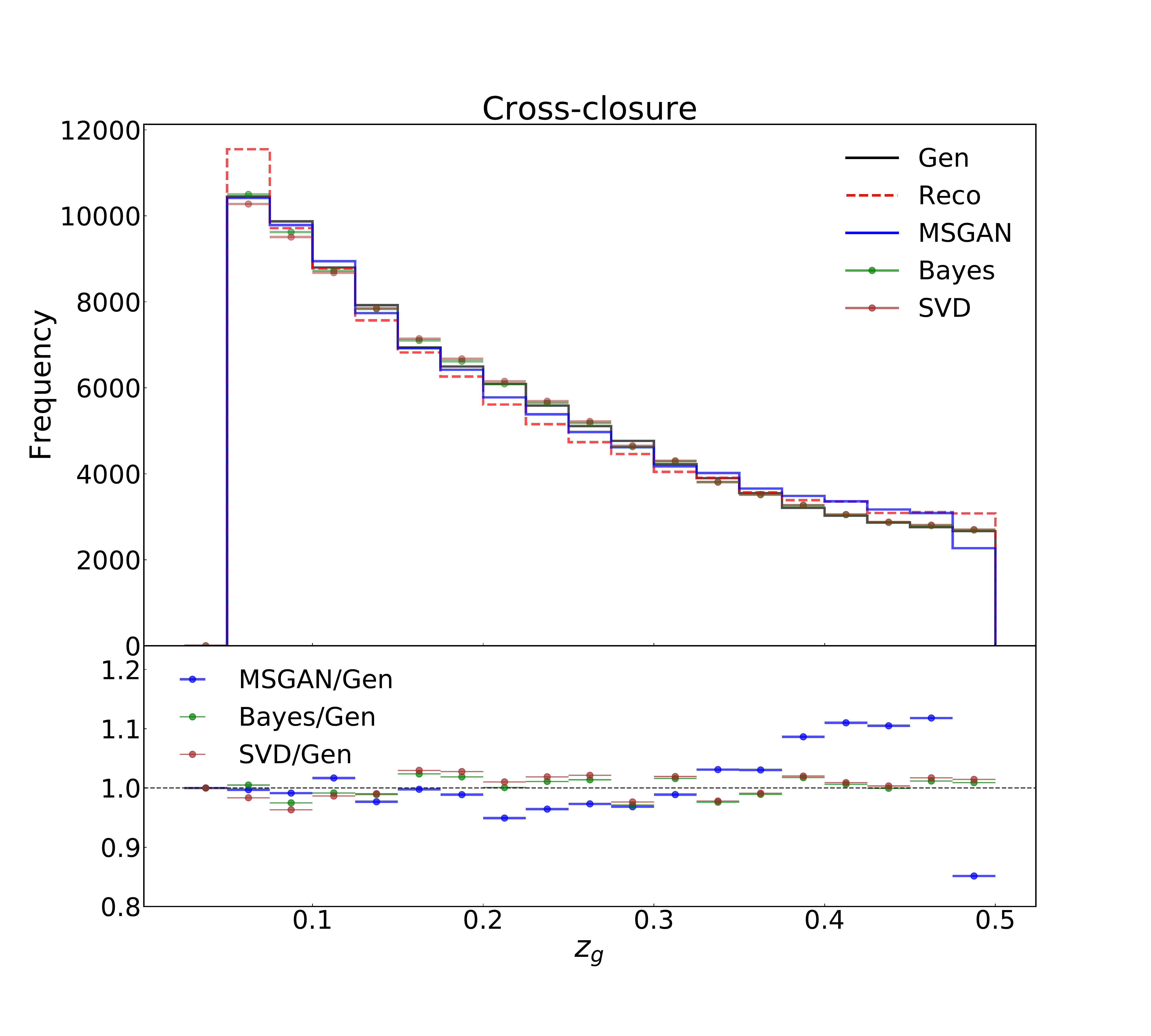}}
	\caption{$Z_{g}$ unfolding}
	\label{fig:zg}
\end{figure*}

\begin{figure*}[!htb]
	\centering
	\subfigure[]{\includegraphics[width=0.48\textwidth]{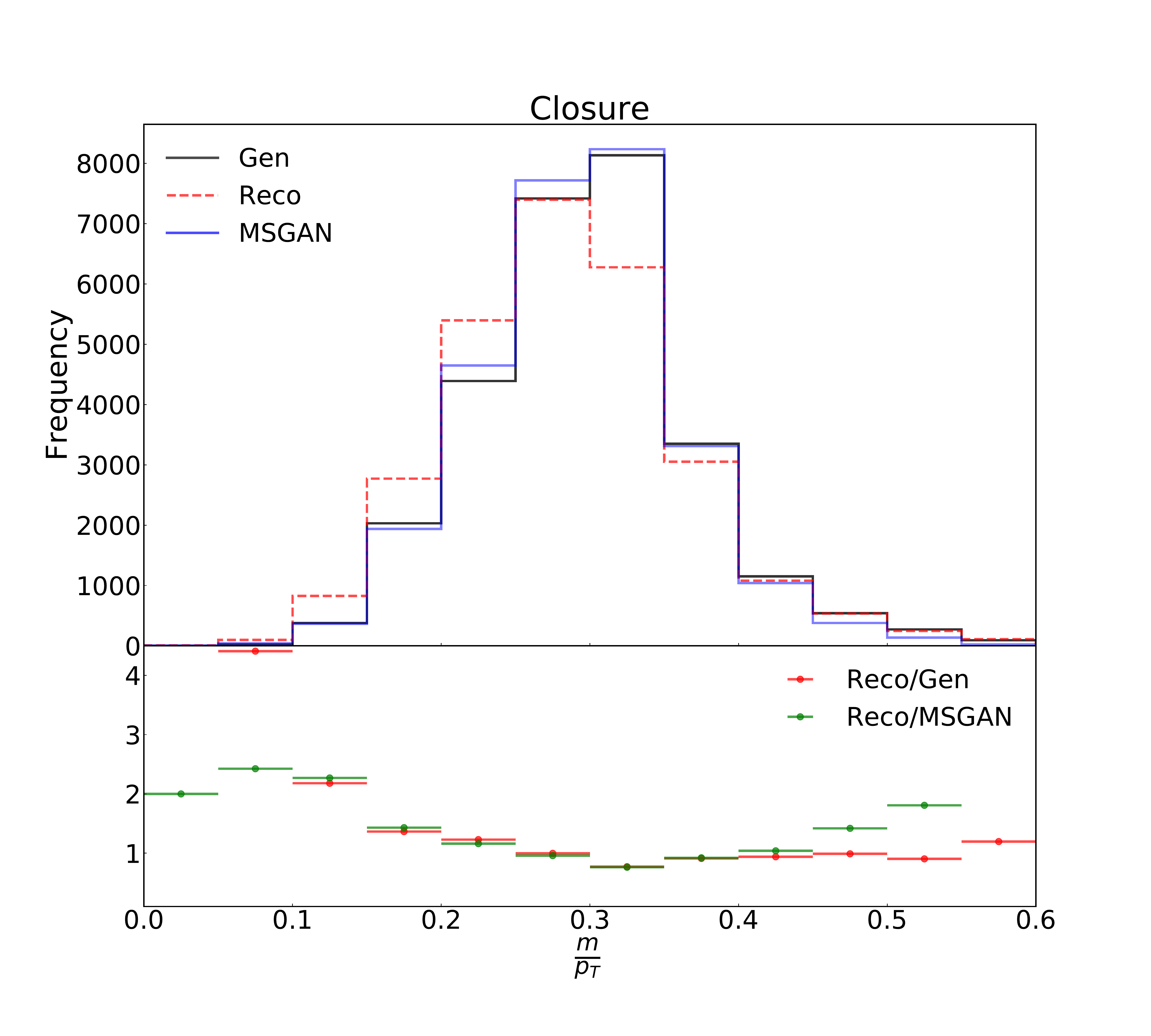}}\quad
	\subfigure[]{\includegraphics[width=0.48\textwidth]{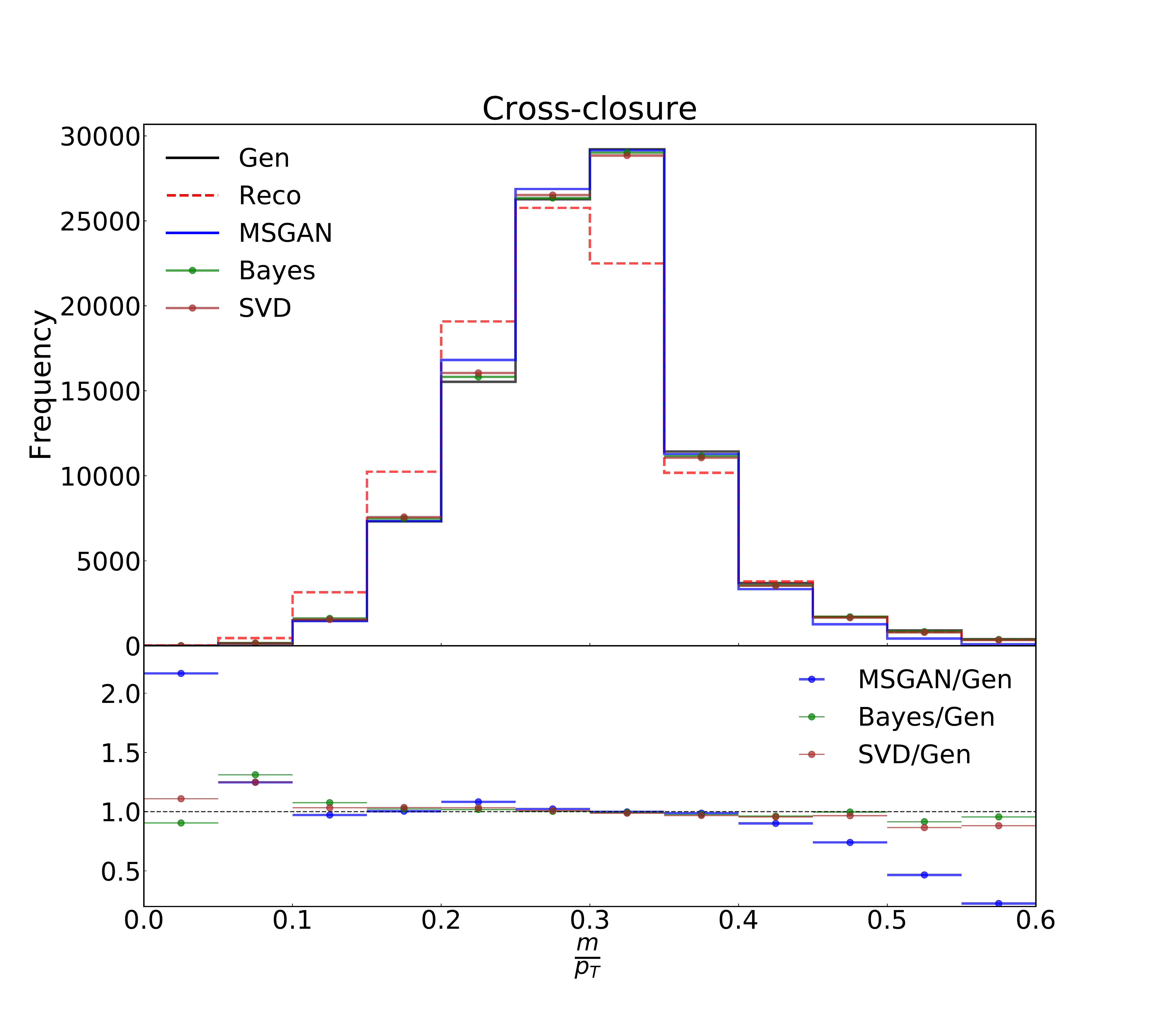}}
	\caption{Normalized mass unfolding}
	\label{fig:nmass}
\end{figure*}

\begin{figure*}[!htb]
	\centering
	\subfigure[]{\includegraphics[width=0.48\textwidth]{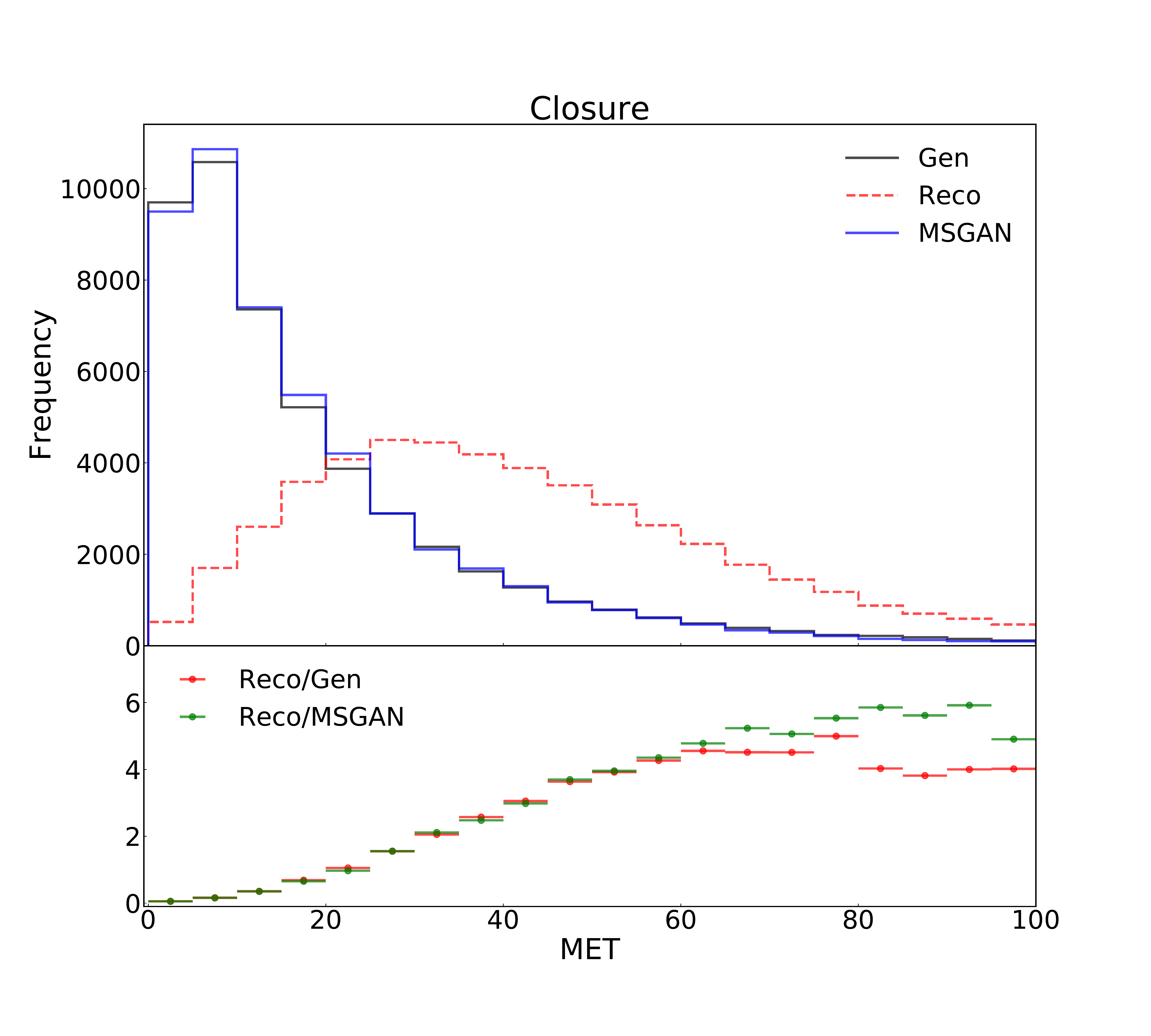}}\quad
	\subfigure[]{\includegraphics[width=0.48\textwidth]{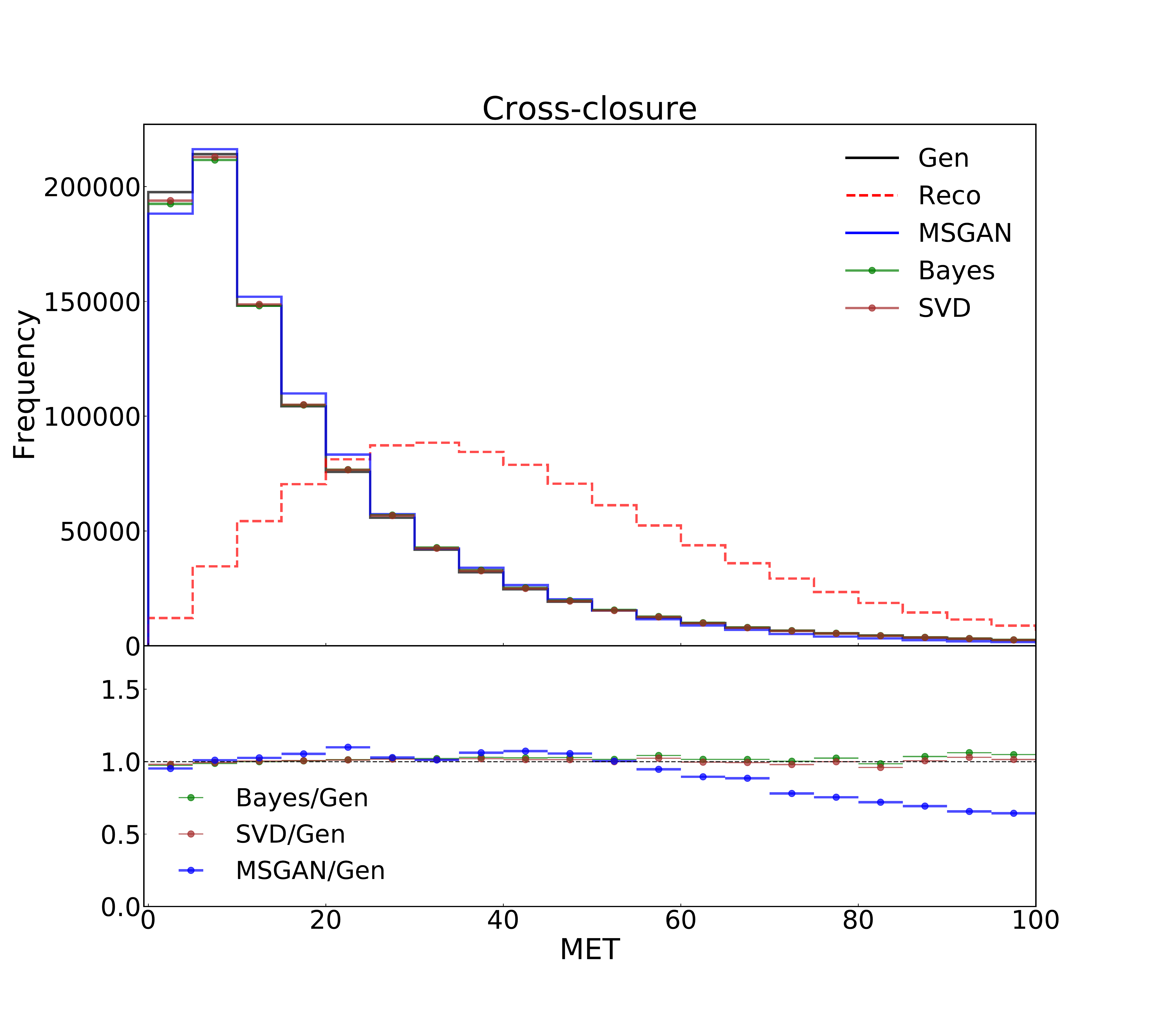}}
	\caption{Missing Transverse Energy unfolding}
	\label{fig:met}
\end{figure*}

The performance of MSGAN unfolding for the four selected JSS observables and MET 
is compared with the Bayesian and SVD results, in Figs.~\ref{fig:ecf} -- \ref{fig:met}.

\FloatBarrier

\FloatBarrier

Two sets of plots are presented for each variable. The left panel (marked \textit{closure})
shows the distributions used to derive the smearing matrices for the standard unfolding methods, 
and train the MSGAN for the approach introduced here. 
The MSGAN lines here are drawn using a subset ($5\%$) of the same samples, which was not used in training.
The ratios, Reco/Gen and Reco/MSGAN will be identical if the MSGAN learned the transform perfectly. 

In the right panel in each case 
(marked \textit{cross-closure}), the same distributions are shown, 
but this time generated with the alternate tune, resulting in somewhat different shapes.
The Bayesian and SVD unfolded lines are derived using the smearing matrices derived from the left panel, 
and the MSGAN line is produced using the already trained algorithm. 
The ratios compare the output of MSGAN with Bayesian and SVD, where values
closer to unity indicate ideal unfolding performance.

In all figures, the trend followed by Reco/MSGAN eventually reflects the trend of Reco/Gen, 
thus indicating that MSGAN is effectively learning the detector effects encoded in mapping from reco to gen. 
In all cases, the MSGAN's performance in the interesting part of the distribution
is seen to be as good as the standard methods. 
However, it falters a bit in the statistically limited tails. The reduced performance 
for $z_g$ and in the tail of MET distribution is probably due to its flat shape over the range 
of input values.

\section{\label{sec:out} Outlook}

In this work, an application of Generative Adversarial Networks is shown for unfolding detector-level distributions to particle-level.
This is a proof-of-principle demonstration that the proposed MSGAN architecture can successfully learn to model detector effects, 
however we are still not at a stage where the approach can replace traditional methods perfected over many decades.

Even when the shapes of the reco and gen distributions are markedly different, the MSGAN approach performs well, especially
for steeply falling distributions where standard unfolding methods are not always optimal. In this application, the generator of the MSGAN essentially learns a
transfer function encoding a convolution of the detector response and the reconstruction algorithm in a bin independent way. The advantage of that is once the MSGAN generator learns
the mapping for a particular variable, it can be used to unfold directly. It does not depend on the subjective criteria of number of iterations
necessary in Bayesian unfolding, and it requires no reweighting of the input distributions often used. 
We note here, that this method is best applied in use-cases where the difference between reco and gen 
samples are correlated to a consistent set of detector effects. 

The MSGAN, as formulated here, might face difficulty learning a transformation
if the mapping is discontinuous. However, concatenating a noise vector to the 
standard MSGAN input would potentially allow one to mitigate this issue. 
An extra noise input might also help in addressing potential over-dependence on the generator 
and/or detector simulations in use cases where that is a concern. 
If necessary, the MC can also be reweighted to better describe data before the networks are trained on them. 

In principle, the MSGAN framework could also be used for the simulation of 
detector effects by simply reversing the workflow and utilizing gen as the latent space to map to reco. 
This work can also be expanded by 
including the effect of multiple proton-proton interactions (i.e.\textit{ pile-up}) in distributions where it has an effect, and we provide results with the MET distribution in Fig.~\ref{fig:met}
as an example of that.
This approach can also be applied to higher-dimensional, analysis-specific sets of observables simultaneously, thereby allowing the GAN to learn a correlation between gen and reco as a whole, 
in a distribution
independent way, which we leave to future work.

\acknowledgments

We thank G\"unther Dissertori,  Andrew Larkoski and Maurizio Pierini for comments 
on the draft. 
DK is supported by National Research Foundation, South Africa in forms of CSUR and Incentive funding. 
DR is supported by NRF grant-holder linked post-doctoral fellowship.

\bibliography{mlu}

\providecommand{\noopsort}[1]{}\providecommand{\singleletter}[1]{#1}%
\providecommand{\href}[2]{#2}\begingroup\raggedright\begin{thebibliography}{10}

\bibitem{Blobel:2002pu}
V.~Blobel, {\it {An Unfolding method for high-energy physics experiments}},  in
  {\em {Advanced Statistical Techniques in Particle Physics. Proceedings,
  Conference, Durham, UK, March 18-22, 2002}}, pp.~258--267, 2002.
\newblock \href{http://arxiv.org/abs/hep-ex/0208022}{{\tt hep-ex/0208022}}.

\bibitem{D'Agostini:1994zf}
G.~D'Agostini, {\it {A Multidimensional unfolding method based on Bayes'
  theorem}},  {\em Nucl. Instrum. Meth.} {\bf A362} (1995) 487--498.

\bibitem{Hocker:1995kb}
A.~Hocker and V.~Kartvelishvili, {\it {SVD approach to data unfolding}},  {\em
  Nucl. Instrum. Meth.} {\bf A372} (1996) 469--481,
  [\href{http://arxiv.org/abs/hep-ph/9509307}{{\tt hep-ph/9509307}}].

\bibitem{2014arXiv1406.2661G}
I.~J. {Goodfellow}, J.~{Pouget-Abadie}, M.~{Mirza}, B.~{Xu}, D.~{Warde-Farley},
  S.~{Ozair}, A.~{Courville}, and Y.~{Bengio}, {\it {Generative Adversarial
  Networks}},  {\em ArXiv e-prints} (June, 2014)
  [\href{http://arxiv.org/abs/1406.2661}{{\tt arXiv:1406.2661}}].

\bibitem{Gagunashvili:2010zw}
N.~D. Gagunashvili, {\it {Machine learning approach to inverse problem and
  unfolding procedure}},  \href{http://arxiv.org/abs/1004.2006}{{\tt
  arXiv:1004.2006}}.

\bibitem{Glazov:2017vni}
A.~Glazov, {\it {Machine learning as an instrument for data unfolding}},
  \href{http://arxiv.org/abs/1712.01814}{{\tt arXiv:1712.01814}}.

\bibitem{Sjostrand:2007gs}
T.~Sj{\"o}strand, S.~Mrenna, and P.~Skands, {\it {A brief introduction to
  PYTHIA 8.1}},  {\em Comput. Phys. Commun.} {\bf 178} (2008) 852--867,
  [\href{http://arxiv.org/abs/0710.3820}{{\tt arXiv:0710.3820}}].

\bibitem{Sjostrand:2014zea}
T.~Sj{\"o}strand, S.~Ask, J.~R. Christiansen, R.~Corke, N.~Desai, P.~Ilten,
  S.~Mrenna, S.~Prestel, C.~O. Rasmussen, and P.~Z. Skands, {\it {An
  Introduction to PYTHIA 8.2}},  {\em Comput. Phys. Commun.} {\bf 191} (2015)
  159--177, [\href{http://arxiv.org/abs/1410.3012}{{\tt arXiv:1410.3012}}].

\bibitem{Skands:2014pea}
P.~Skands, S.~Carrazza, and J.~Rojo, {\it {Tuning PYTHIA 8.1: the Monash 2013
  Tune}},  {\em Eur. Phys. J.} {\bf C74} (2014), no.~8 3024,
  [\href{http://arxiv.org/abs/1404.5630}{{\tt arXiv:1404.5630}}].

\bibitem{Fischer:2016zzs}
N.~Fischer and T.~Sjöstrand, {\it {Thermodynamical String Fragmentation}},
  {\em JHEP} {\bf 01} (2017) 140, [\href{http://arxiv.org/abs/1610.09818}{{\tt
  arXiv:1610.09818}}].

\bibitem{Buckley:2010ar}
A.~Buckley, J.~Butterworth, L.~Lonnblad, D.~Grellscheid, H.~Hoeth, J.~Monk,
  H.~Schulz, and F.~Siegert, {\it {Rivet user manual}},  {\em Comput. Phys.
  Commun.} {\bf 184} (2013) 2803--2819,
  [\href{http://arxiv.org/abs/1003.0694}{{\tt arXiv:1003.0694}}].

\bibitem{Catani:1993hr}
S.~Catani, Y.~L. Dokshitzer, M.~H. Seymour, and B.~R. Webber, {\it
  {Longitudinally invariant $k_\perp$ clustering algorithms for hadron hadron
  collisions}},  {\em Nucl. Phys. B} {\bf 406} (1993) 187--224.

\bibitem{Larkoski:2013eya}
A.~J. Larkoski, G.~P. Salam, and J.~Thaler, {\it {Energy Correlation Functions
  for Jet Substructure}},  {\em JHEP} {\bf 06} (2013) 108,
  [\href{http://arxiv.org/abs/1305.0007}{{\tt arXiv:1305.0007}}].

\bibitem{Gras:2017jty}
P.~Gras, S.~Höche, D.~Kar, A.~Larkoski, L.~Lönnblad, S.~Plätzer,
  A.~Siódmok, P.~Skands, G.~Soyez, and J.~Thaler, {\it {Systematics of
  quark/gluon tagging}},  {\em JHEP} {\bf 07} (2017) 091,
  [\href{http://arxiv.org/abs/1704.03878}{{\tt arXiv:1704.03878}}].

\bibitem{Larkoski:2014wba}
A.~J. Larkoski, S.~Marzani, G.~Soyez, and J.~Thaler, {\it {Soft Drop}},  {\em
  JHEP} {\bf 05} (2014) 146, [\href{http://arxiv.org/abs/1402.2657}{{\tt
  arXiv:1402.2657}}].

\bibitem{Alwall:2014hca}
J.~Alwall, R.~Frederix, S.~Frixione, V.~Hirschi, F.~Maltoni, O.~Mattelaer,
  H.~S. Shao, T.~Stelzer, P.~Torrielli, and M.~Zaro, {\it {The automated
  computation of tree-level and next-to-leading order differential cross
  sections, and their matching to parton shower simulations}},  {\em JHEP} {\bf
  07} (2014) 079, [\href{http://arxiv.org/abs/1405.0301}{{\tt
  arXiv:1405.0301}}].

\bibitem{deFavereau:2013fsa}
{\bf DELPHES 3} Collaboration, J.~de~Favereau, C.~Delaere, P.~Demin,
  A.~Giammanco, V.~Lemaître, A.~Mertens, and M.~Selvaggi, {\it {DELPHES 3, A
  modular framework for fast simulation of a generic collider experiment}},
  {\em JHEP} {\bf 02} (2014) 057, [\href{http://arxiv.org/abs/1307.6346}{{\tt
  arXiv:1307.6346}}].

\bibitem{deOliveira:2017pjk}
L.~de~Oliveira, M.~Paganini, and B.~Nachman, {\it {Learning Particle Physics by
  Example: Location-Aware Generative Adversarial Networks for Physics
  Synthesis}},  \href{http://arxiv.org/abs/1701.05927}{{\tt arXiv:1701.05927}}.

\bibitem{deOliveira:2017rwa}
L.~de~Oliveira, M.~Paganini, and B.~Nachman, {\it {Controlling Physical
  Attributes in GAN-Accelerated Simulation of Electromagnetic Calorimeters}},
  in {\em {18th International Workshop on Advanced Computing and Analysis
  Techniques in Physics Research (ACAT 2017) Seattle, WA, USA, August 21-25,
  2017}}, 2017.
\newblock \href{http://arxiv.org/abs/1711.08813}{{\tt arXiv:1711.08813}}.

\bibitem{Paganini:2017dwg}
M.~Paganini, L.~de~Oliveira, and B.~Nachman, {\it {CaloGAN : Simulating 3D high
  energy particle showers in multilayer electromagnetic calorimeters with
  generative adversarial networks}},  {\em Phys. Rev.} {\bf D97} (2018), no.~1
  014021, [\href{http://arxiv.org/abs/1712.10321}{{\tt arXiv:1712.10321}}].

\bibitem{Paganini:2017hrr}
M.~Paganini, L.~de~Oliveira, and B.~Nachman, {\it {Accelerating Science with
  Generative Adversarial Networks: An Application to 3D Particle Showers in
  Multilayer Calorimeters}},  {\em Phys. Rev. Lett.} {\bf 120} (2018), no.~4
  042003, [\href{http://arxiv.org/abs/1705.02355}{{\tt arXiv:1705.02355}}].

\bibitem{Musella:2018rdi}
P.~Musella and F.~Pandolfi, {\it {Fast and accurate simulation of particle
  detectors using generative adversarial neural networks}},
  \href{http://arxiv.org/abs/1805.00850}{{\tt arXiv:1805.00850}}.

\bibitem{NIPS2016_6125}
T.~Salimans, I.~Goodfellow, W.~Zaremba, V.~Cheung, A.~Radford, X.~Chen, and
  X.~Chen, {\it Improved techniques for training gans},  in {\em Advances in
  Neural Information Processing Systems 29} (D.~D. Lee, M.~Sugiyama, U.~V.
  Luxburg, I.~Guyon, and R.~Garnett, eds.), pp.~2234--2242.
\newblock Curran Associates, Inc., 2016.

\bibitem{8190958}
H.~Y. Lee, J.~M. Kwak, B.~Ban, S.~J. Na, S.~R. Lee, and H.~K. Lee, {\it Gan-d:
  Generative adversarial networks for image deconvolution},  in {\em 2017
  International Conference on Information and Communication Technology
  Convergence (ICTC)}, pp.~132--137, Oct, 2017.

\bibitem{Maas13rectifiernonlinearities}
A.~L. Maas, A.~Y. Hannun, and A.~Y. Ng, {\it Rectifier nonlinearities improve
  neural network acoustic models},  in {\em in ICML Workshop on Deep Learning
  for Audio, Speech and Language Processing}, 2013.

\bibitem{DBLP:journals/corr/KingmaB14}
D.~P. Kingma and J.~Ba, {\it Adam: {A} method for stochastic optimization},
  {\em CoRR} {\bf abs/1412.6980} (2014)
  [\href{http://arxiv.org/abs/1412.6980}{{\tt arXiv:1412.6980}}].

\bibitem{2016arXiv160603498S}
T.~{Salimans}, I.~{Goodfellow}, W.~{Zaremba}, V.~{Cheung}, A.~{Radford}, and
  X.~{Chen}, {\it {Improved Techniques for Training GANs}},  {\em ArXiv
  e-prints} (June, 2016) [\href{http://arxiv.org/abs/1606.03498}{{\tt
  arXiv:1606.03498}}].

\bibitem{chollet2015keras}
F.~Chollet, ``Keras.'' \url{https://github.com/fchollet/keras}, 2015.

\bibitem{adye_2011}
T.~Adye, ``Unfolding algorithms and tests using {RooUnfold}.'' Proceedings of
  the {PHYSTAT} 2011 Workshop, {CERN, Geneva, Switzerland}, January 2011,
  CERN-2011-006, pp 313-318, 2011.

\end{thebibliography}\endgroup
\end{document}